\documentclass{svmult}

\usepackage{makeidx}         % allows index generation
\usepackage{graphicx}        % standard LaTeX graphics tool
\usepackage{multicol}        % used for the two-column index
\usepackage[bottom]{footmisc}% places footnotes at page bottom

\usepackage{float}
\usepackage{subfigure}
\usepackage{multirow}
\usepackage{algorithm} %format of the algorithm
\usepackage[noend]{algorithmic} %format of the algorithm
\usepackage{amsmath}
\usepackage{amssymb}
\usepackage{cite}
\makeindex             % used for the subject index
\begin{document}

\title*{The Modified Direct Method: an Approach for Smoothing Planar and Surface Meshes}
\titlerunning{The Modified Direct Method(MDM)}
\author{
Gang Mei\inst{1}\,
John C.Tipper\inst{1}\and
Nengxiong Xu\inst{2}}
\institute{Institut f\"ur Geowissenschaften-Geologie, Albert-Ludwigs-Universit\"at Freiburg, Albertstra\ss e 23B, D-79104, Freiburg im Breisgau, Germany.
\texttt{\{gang.mei,john.tipper\}@geologie.uni-freiburg.de}
\and School of Engineering and Technology, China University of Geosciences, Beijing, 100083, China.
\texttt{xunengxiong@yahoo.com.cn}}
\maketitle

\begin{abstract}
The Modified Direct Method (MDM) is an iterative mesh smoothing method for smoothing planar and surface meshes, which is developed from the non-iterative smoothing method originated by Balendran\cite{1.}. When smooth planar meshes, the performance of the MDM is effectively identical to that of Laplacian smoothing, for triangular and quadrilateral meshes; however, the MDM outperforms Laplacian smoothing for tri-quad meshes. When smooth surface meshes, for  triangular, quadrilateral and quad-dominant mixed meshes, the mean quality(MQ) of all mesh elements always increases and the mean square error (MSE) decreases during smoothing; For tri-dominant mixed mesh, the quality of triangles always descends while that of quads ascends. Test examples show that the MDM is convergent for both planar and surface triangular, quadrilateral and tri-quad meshes.

\end{abstract}

\section{Introduction}
\label{sec:1}
In finite element analysis it is important always to use high quality meshes: low quality meshes lead to unreliable results. A mesh that has been newly created usually needs to be improved before it can be used. This improvement can be made using either (1) mesh clear-up methods, which insert or delete nodes as well as change the connectivity of the mesh elements, or (2) mesh smoothing methods, which leave the element connectivity unchanged and instead reposition the mesh nodes. This paper is mainly concerned with smoothing methods.

There are numerous papers published concerning the topic of mesh smoothing. In this paper, we just refer some that are the most popular or representative of the state of the art for planar and surface meshes.

\subsection{Related works}
\label{sec:1.1:Related works}
Mesh smoothing methods probably can be classified into four types: the geometry-based, the optimization-based, the physics-based and the combined.
The geometry-based methods obtain new location of nodes by geometric rules, local optimization techniques or minimizing objective functions.

The most popular geometry-based smoothing methods is the Laplacian smoothing \cite{13.}, which repositions each node at the centroid of its neighboring nodes in one iteration. The popularity of this method comes from its simplicity and effectiveness. To improve the basic form of Laplacian smoothing, some smart, constrained or weighted variations have been proposed \cite{9., 12., 29.}.

Another simpler but more effective method is called angle-based approach \cite{30.}, in which new nodal locations are calculated by conforming specific angle ratios in the surrounding polygons.
A geometric element transformation method (GETMe) \cite{27.} based on a simple geometric transformation can be applicable to elements bounded by polygons.

A projecting/smoothing method is performed for smoothing surface mesh by minimizing the mean ratio of all triangles sharing the free node\cite{8.}.
A novel method based on quadric surface fitting, vertex projecting, curvature estimating and mesh labeling is applied  in biomedical modeling\cite{28.} .

An effective variational method for smoothing surface and volume triangulations is proposed by Jiao X et al \cite{16.}, where the discrepancies between actual and target elements is reduced by minimizing two energy functions.
Also, a general-purpose algorithm called the target-matrix paradigm is introduced in \cite{18.}, and can be applied to a wide variety of meshes.

Different from the geometry-based methods, in optimization-based ones, the smoothed position of all nodes is acquired by minimizing a given distortion metric. This series of methods is more expensive but can generate better results than most geometry-based ones especially at concave regions. Some literatures devoted on this topic include \cite{3., 11., 22., 24.}.

The physics-based smoothing methods are the techniques that smooth meshes based on physical processing\cite{21.}, or by solving simple physics problems.
Shimada \cite{23.} proposed a method which treats nodes as the center of bubbles and nodal locations are obtained by deforming bubbles with each other. A similar algorithm called pliant method is presented in \cite{2.}.

To improve performance, two or more basic methods can be combined into a hybrid approach. Some hybrid methods are the combination of Laplacian smoothing with optimization-based methods \cite{4., 5., 10.}.

In the aspect of computation, mesh smoothing methods can be either iterative or non-iterative. Most of them are iterative algorithms; a non-iterative method is proposed by Balendran \cite{1.} which is referred to here as the Direct Method (DM). The goal of the DM is simple -- to make triangular elements as close to equilateral as possible and quadrilateral elements as close to square as possible -- and it achieves this goal by generating and solving a set of optimization equations.

\subsection{Our contribution}
\label{sec:1.2:Our contribution}
In this paper we introduce a smoothing method that has the same basic goal as the DM, but is iterative rather than non-iterative; we term this method the Modified Direct Method (MDM) which can be used to smooth both planar and surface meshes.

The main procedure of MDM for smoothing planar meshes is relatively simple: Firstly element stiffness matrices are created based on the type of elements. The modified forms of element stiffness matrices are simpler than those of DM. And then by assembling all element stiffness matrices, a system of Jacobi iteration equations can be formed, which is different from the optimization equations in DM. Finally, the smoothed nodal coordinates can be generated by solving the system of Jacobi iteration equations.

For smoothing surface meshes, the MDM becomes complex:
Firstly to maintain features of original meshes, feature points are detected and then fixed as constrained nodes. Then a Jacobi iteration matrix which is similar to that in 2D is also assembled. Thirdly the relocated position of every node in each iteration is calculated according to the Jacobi iteration matrix, and then projected onto the original mesh to be smoothed position in current iteration. And finally, the required smoothed nodal coordinates can be obtained until the quality of smoothed mesh will not improve.

The paper is organized as follows. In Sect.2, we first give a brief description of the basics of the original DM. Then in Sect. 3, we show how the DM is modified and developed into the MDM for smoothing planar and surface meshes; also we introduce several key techniques such as features detection of the MDM for smoothing surface meshes. Finally we present some tests of the MDM and make a convergence analysis of the MDM for smoothing surface meshes, and summarize them in Sect.4.

%2 The Direct Method (DM)
\section{The Direct Method (DM)}
\label{sec:2}

The optimization equations solved in the DM are generated by assembling
element stiffness matrices into a global stiffness matrix. These element
stiffness matrices are 6$\times $6 in size for planar triangular elements
and 8$\times $8 in size for planar quadrilateral elements. The global
stiffness matrix for a mesh with $n$ nodes is $2n\times 2n$ in size,
irrespective of whether the elements are triangles or quadrilaterals.

\subsection{Planar triangular mesh}  \label{sec:2.1}

\subsubsection{The element stiffness matrix}

Consider a triangular element ABC shown in Fig. 1. A$^{\ast }$ is
the position to which node A would have to be moved to make the element
equilateral, assuming that nodes B and C were fixed; B$^{\ast }$ is the
position to which B would have to be moved assuming A and C were fixed;
C$^{\ast }$ is the position to which C would have to be moved assuming A and
B were fixed. The coordinates of A$^{\ast }$ are:

\begin{equation}
\label{eq1}
\left\{ {{\begin{array}{*{20}l}
 {X_A^* = \dfrac{1}{2}(X_B +X_C )+\dfrac{\sqrt 3 }{2}(Y_B -Y_C )} \\
 {Y_A^* = \dfrac{\sqrt 3 }{2}(-X_B +X_C )+\dfrac{1}{2}(Y_B +Y_C )}
\\
\end{array} }} \right.
\end{equation}

\begin{figure}[H]
\centering
\includegraphics[height = 45mm]{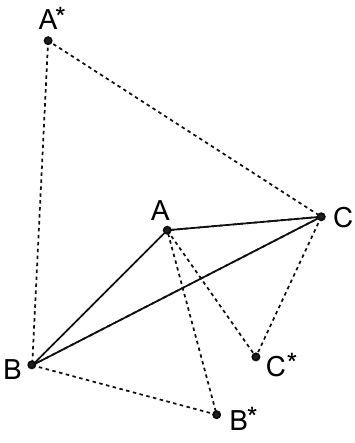}
\caption{Triangular element ABC.  A*, B*, C* are positions to which nodes would have to be moved singly in order to make ABC equilateral. See Eq. \ref{eq3} }
\label{fig:1:DM:T}       % Give a unique label
\end{figure}

These equations can be rewritten as:
\begin{equation}
\label{eq2}
\left\{ {{\begin{array}{*{20}l}
 {2(X_A -X_A^* )=(2X_A +0Y_A )-(X_B +\sqrt 3 Y_B )+(-X_C +\sqrt 3 Y_C )}
\\
 {2(Y_A -Y_A^* )=(0X_A +2Y_A )+(\sqrt 3 X_B -Y_B )-(\sqrt 3 X_C +Y_C )}
\\
\end{array} }} \right.
\end{equation}
The coordinates of B$^{\ast }$ and C$^{\ast }$ are obtained in the same way.
The coordinates of A, B, C, A$^{\ast }$, B$^{\ast }$ and C$^{\ast }$ are
related by the following equations:
\begin{equation}\label{eq3}
\begin{bmatrix}
2 & 0 & {-1} & {-\sqrt 3 } & {-1} & {\sqrt 3 } \\
0 & 2 & {\sqrt 3 } & {-1} &{-\sqrt 3 }& {-1} \\
{-1} & {\sqrt 3 } & 2 & 0 & {-1} & {-\sqrt 3 } \\
{-\sqrt 3 } & {-1} & 0 & 2 & {\sqrt 3 } & {-1} \\
{-1} & {-\sqrt 3 } & {-1} & {\sqrt 3 } & 2 & 0 \\
{\sqrt 3 } & {-1} & {-\sqrt 3 } & {-1} & 0 & 2
\end{bmatrix}
\begin{bmatrix}
X_A \\
Y_A \\
X_B \\
Y_B \\
X_C \\
Y_C
\end{bmatrix}
=
\begin{bmatrix}
2(X_A-X_A^*) \\
2(Y_A-Y_A^*) \\
2(X_B-X_B^*) \\
2(Y_B-Y_B^*) \\
2(X_C-X_C^*) \\
2(Y_C-Y_C^*)
\end{bmatrix}
\end{equation}

The left-hand matrix in Eq. \ref{eq3} is termed the stiffness matrix of the planar
triangular element.

\subsubsection{The global stiffness matrix and the optimization equations}

\begin{figure}[H]
\centering
\includegraphics[height = 37mm]{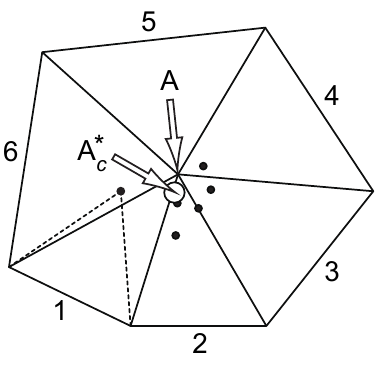}
\caption{Node A belongs to 6 triangular elements. A* can be calculated for each element separately (black circles). Optimal smoothed position for A is the centroid of these, $A_c^*$(arrowed open circle)}
\label{fig:2:DM}       % Give a unique label
\end{figure}

Now assume that ABC is part of a planar triangular mesh that has $n$ nodes.
Each node of ABC -- for instance node A -- is then shared with several other
elements, and A$^{\ast }$ can be calculated for each of these. The final
position of A -- its optimal smoothed position -- is obtained by averaging
the separately calculated A$^{\ast }$'s (Fig. \ref{fig:2:DM}). (This averaging process is
effectively identical to that used in Laplacian smoothing, which explains
why the test results obtained using Laplacian smoothing are identical to
those obtained using MDM, for uniformly triangular and uniformly
quadrilateral meshes -- see Figs \ref{fig:4:Smooth:T}, \ref{fig:5:Smooth:Q}.)

The smoothed positions for the complete set of $n$ nodes are given by the
equations:
\begin{eqnarray}\label{eq4}
\begin{bmatrix}
 {\alpha _{11} } & {\alpha _{12} } & {\alpha _{13} } & \cdots   & {\alpha _{1(2n)} } \\
 {\alpha _{21} }  & {\alpha _{22} }  & {\alpha _{23} }  & \cdots  & {\alpha _{2(2n)} }  \\
 {\alpha _{31} }  & {\alpha _{32} }  & {\alpha _{33} }  & \cdots  & {\alpha _{3(2n)} }  \\
 \vdots  & \vdots  & \vdots   & \ddots &   \vdots  \\
 {\alpha _{(2n)1} }  & {\alpha _{(2n)2} }  & {\alpha _{(2n)3} }
 & \cdots   & {\alpha _{(2n)(2n)} }  \\
\end{bmatrix}
\begin{bmatrix}
 {X_1 }  \\
 {Y_1 }  \\
 {X_2 }  \\
 \vdots  \\
 {Y_n }
\end{bmatrix}
=\begin{bmatrix}
 {e_1\cdot 2(X_1 -X_1^\ast )}  \\
 {e_1 \cdot 2(Y_1 -Y_1^\ast )}  \\
 {e_2 \cdot 2(X_2 -X_2^\ast )}  \\
 \vdots  \\
 {e_n \cdot 2(Y_n -Y_n^\ast )}
\end{bmatrix}
\end{eqnarray}

These equations are the optimization equations that need to be solved in the
DM. The left-hand matrix -- this is termed the global stiffness matrix -- is
created by assembling the individual element stiffness matrices according to
the connectivity of the elements in the mesh concerned. The elements in the
global stiffness matrix are obtained during this assembly process, and they
will of course be different from mesh to mesh. $e_i (1\le i\le n)$is the
number of elements in the mesh that share node $i$.

\subsection{Planar quadrilateral mesh}  \label{sec:2.2}

\subsubsection{The element stiffness matrix}

Consider next a quadrilateral element ABCD shown in Fig. \ref{eq3}.
A$^{\ast }$ and C$^{\ast }$ are the positions to which nodes A and C would
have to be moved to make the element square, assuming B and D
were fixed. The coordinates of A$^{\ast }$ are:

\begin{figure}[H]
\centering
\includegraphics[height = 31mm]{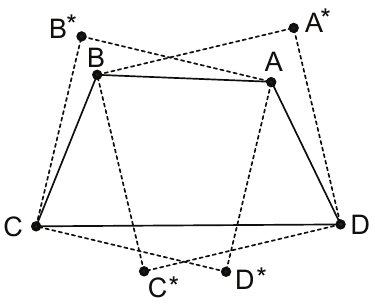}
\caption{Quadrilateral element ABCD. A*, B*, C*, D* are positions to which diagonally opposite nodes would have to be moved to make ABCD square. See Eq.\ref{eq7}}
\label{fig:3:DM:Q}       % Give a unique label
\end{figure}

\begin{equation}
\label{eq5}
\left\{ {{\begin{array}{*{20}l}
 {X_A^* = \dfrac{1}{2}(X_B +X_D )+\dfrac{1}{2}(Y_B -Y_D )} \\
 {Y_A^* = \dfrac{1}{2}(-X_B +X_D )+\dfrac{1}{2}(Y_B +Y_D )} \\
\end{array} }} \right.
\end{equation}

Similarly as that to triangular element, the coordinates of A, B, C, D, A$^{\ast }$, B$^{\ast }$,
C$^{\ast }$ and D$^{\ast }$ are related by the following equations:
\begin{equation}
\label{eq7}
\begin{bmatrix}
 2  & 0  & {-1}  & {-1}  & 0  & 0  &
{-1}  & 1  \\
 0  & 2  & 1  & {-1}  & 0  & 0  & {-1}
 & {-1}  \\
 {-1}  & 1  & 2  & 0  & {-1}  & {-1}  &
0  & 0  \\
 {-1}  & {-1}  & 0  & 2  & 1  & {-1}  &
0  & 0  \\
 0  & 0  & {-1}  & 1  & 2  & 0  & {-1}
 & {-1}  \\
 0  & 0  & {-1}  & {-1}  & 0  & 2  & 1
 & {-1}  \\
 {-1}  & {-1}  & 0  & 0  & {-1}  & 1  &
2  & 0  \\
 1  & {-1}  & 0  & 0  & {-1}  & {-1}  &
0  & 2  \\
\end{bmatrix}
\begin{bmatrix}
 {X_A }  \\
 {Y_A }  \\
 {X_B }  \\
 {Y_B }  \\
 {X_C }  \\
 {Y_C }  \\
 {X_D }  \\
 {Y_D }  \\
\end{bmatrix}
=\begin{bmatrix}
 {2(X_A -X_A^\ast )}  \\
 {2(Y_A -Y_A^\ast )}  \\
 {2(X_B -X_B^\ast )}  \\
 {2(Y_B -Y_B^\ast )}  \\
 {2(X_C -X_C^\ast )}  \\
 {2(Y_C -Y_C^\ast )}  \\
 {2(X_D -X_D^\ast )}  \\
 {2(Y_D -Y_D^\ast )}  \\
\end{bmatrix}
\end{equation}
The left-hand matrix in Eq. \ref{eq7} is termed the stiffness matrix of the planar
quadrilateral element.

\subsubsection{The global stiffness matrix and the optimization equations}

Now assume ABCD is part of a quadrilateral mesh that has $n$ nodes. As before,
the global stiffness matrix is obtained by assembling the element stiffness
matrices according to the connectivity of the elements in the mesh
concerned. The optimization equations associated with this global stiffness
matrix are identical in form to those for the triangular mesh (Eq. \ref{eq4}).

% 3 The Modified Direct Method (MDM)
\section{The Modified Direct Method (MDM)}
\label{sec:3}

The development of the MDM from the DM involves (\ref{eq1}) the use of different
element stiffness matrices, (\ref{eq2}) the use of a Jacobian iteration matrix
instead of a global stiffness matrix, and (\ref{eq3}) the replacement of the
optimization equations with iteration equations. The mathematical steps
involved in this development are broadly similar for the triangular mesh,
the quadrilateral mesh and the tri-quad mesh.

\subsection{The element stiffness matrices }  \label{sec:3.1}
\label{sec:3.1}

\subsubsection{Triangular mesh }

According to Eq. 1, the coordinates of A, B, C, A$^{\ast }$, B$^{\ast }$ and C$^{\ast }$ are
then related by the following equations:
\begin{equation}
\label{eq9}
\begin{bmatrix}
 0  & 0  & {1/2}  & {\sqrt 3 /2}  & {1/2}  &
{-\sqrt 3 /2}  \\
 0  & 0  & {-\sqrt 3 /2}  & {1/2}  & {\sqrt 3 /2}
 & {1/2}  \\
 {1/2}  & {-\sqrt 3 /2}  & 0  & 0  & {1/2}  &
{\sqrt 3 /2}  \\
 {\sqrt 3 /2}  & {1/2}  & 0  & 0  & {-\sqrt 3 /2}
 & {1/2}  \\
 {1/2}  & {\sqrt 3 /2}  & {1/2}  & {-\sqrt 3 /2}  &
0  & 0  \\
 {-\sqrt 3 /2}  & {1/2}  & {\sqrt 3 /2}  & {1/2}  &
0  & 0  \\
\end{bmatrix}
\begin{bmatrix}
 {X_A }  \\
 {Y_A }  \\
 {X_B }  \\
 {Y_B }  \\
 {X_C }  \\
 {Y_C }  \\
\end{bmatrix}
=\begin{bmatrix}
 {X_A^\ast }  \\
 {Y_A^\ast }  \\
 {X_B^\ast }  \\
 {Y_B^\ast }  \\
 {X_C^\ast }  \\
 {Y_C^\ast }  \\
\end{bmatrix}
\end{equation}
These equations can also be rewritten as iteration equations:
\begin{equation}
\label{eq10}
\begin{bmatrix}
 0  & 0  & {1/2}  & {\sqrt 3 /2}  & {1/2}  &
{-\sqrt 3 /2}  \\
 0  & 0  & {-\sqrt 3 /2}  & {1/2}  & {\sqrt 3 /2}
 & {1/2}  \\
 {1/2}  & {-\sqrt 3 /2}  & 0  & 0  & {1/2}  &
{\sqrt 3 /2}  \\
 {\sqrt 3 /2}  & {1/2}  & 0  & 0  & {-\sqrt 3 /2}
 & {1/2}  \\
 {1/2}  & {\sqrt 3 /2}  & {1/2}  & {-\sqrt 3 /2}  &
0  & 0  \\
 {-\sqrt 3 /2}  & {1/2}  & {\sqrt 3 /2}  & {1/2}  &
0  & 0
\end{bmatrix}
\begin{bmatrix}
 {X_A^k }  \\
 {Y_A^k }  \\
 {X_B^k }  \\
 {Y_B^k }  \\
 {X_C^k }  \\
 {Y_C^k }  \\
\end{bmatrix}
=\begin{bmatrix}
 {X_A^{k+1} }  \\
 {Y_A^{k+1} }  \\
 {X_B^{k+1} }  \\
 {Y_B^{k+1} }  \\
 {X_C^{k+1} }  \\
 {Y_C^{k+1} }  \\
\end{bmatrix}
\end{equation}
$X_A^k $ and $Y_A^k $ are the coordinates of node A at step $k$, and $X_A^{k+1} $
and $Y_A^{k+1} $ are the coordinates at step $k$+1; the notation for nodes B
and C is similar. The left-hand matrix in Eq.\ref{eq10} is also the element
stiffness matrix for a planar triangular element, but it is simpler in form
to that used in the DM (Eq.\ref{eq3}).

In Cartesian coordinates, the coordinates of a vertex x, y and z are
equivalent and can be cycled in sequence: $X\to Y,Y\to Z,Z\to X$. Hence, we
easily extend the stiffness matrix (Eq.\ref{eq10}) for a planar triangular
element to a surface one in 3D (Eq.\ref{eq11}).

\begin{equation}
\label{eq11}
\begin{bmatrix}
 0  & 0  & 0  & {1/2}  & {\sqrt 3 /2}  &
{-\sqrt 3 /2}  & {1/2}  & {-\sqrt 3 /2}  & {\sqrt 3 /2}
 \\
 0  & 0  & 0  & {-\sqrt 3 /2}  & {1/2}  &
{\sqrt 3 /2}  & {\sqrt 3 /2}  & {1/2}  & {-\sqrt 3 /2}
 \\
 0  & 0  & 0  & {\sqrt 3 /2}  & {-\sqrt 3 /2}
& {1/2}  & {-\sqrt 3 /2}  & {\sqrt 3 /2}  & {1/2}
\\
 {1/2}  & {-\sqrt 3 /2}  & {\sqrt 3 /2}  & 0  & 0
 & 0  & {1/2}  & {\sqrt 3 /2}  & {-\sqrt 3 /2}
 \\
 {\sqrt 3 /2}  & {1/2}  & {-\sqrt 3 /2}  & 0  & 0
 & 0  & {-\sqrt 3 /2}  & {1/2}  & {\sqrt 3 /2}
 \\
 {-\sqrt 3 /2}  & {\sqrt 3 /2}  & {1/2}  & 0  & 0
 & 0  & {\sqrt 3 /2}  & {-\sqrt 3 /2}  & {1/2}
 \\
 {1/2}  & {\sqrt 3 /2}  & {-\sqrt 3 /2}  & {1/2}  &
{-\sqrt 3 /2}  & {\sqrt 3 /2}  & 0  & 0  & 0
\\
 {-\sqrt 3 /2}  & {1/2}  & {\sqrt 3 /2}  & {\sqrt 3 /2}
 & {1/2}  & {-\sqrt 3 /2}  & 0  & 0  & 0
 \\
 {\sqrt 3 /2}  & {-\sqrt 3 /2}  & {1/2}  & {-\sqrt 3 /2}
 & {\sqrt 3 /2}  & {1/2}  & 0  & 0  & 0
\\
\end{bmatrix}
\end{equation}

\subsubsection{Quadrilateral mesh }

According to Eq.\ref{eq5}, The coordinates of A, B, C, D, A$^{\ast }$, B$^{\ast }$, C$^{\ast }$ and
D$^{\ast }$ can also be rewritten as iteration equations:
\begin{equation}
\label{eq14}
\begin{bmatrix}
 0   & 0   & {1/2}   & {1/2}   & 0   & 0   &
{1/2}   & {-1/2}   \\
 0   & 0   & {-1/2}   & {1/2}   & 0   & 0   &
{1/2}   & {1/2}   \\
 {1/2}   & {-1/2}   & 0   & 0   & {1/2}   & {1/2}
  & 0   & 0   \\
 {1/2}   & {1/2}   & 0   & 0   & {-1/2}   & {1/2}
  & 0   & 0   \\
 0   & 0   & {1/2}   & {-1/2}   & 0   & 0   &
{1/2}   & {1/2}   \\
 0   & 0   & {1/2}   & {1/2}   & 0   & 0   &
{-1/2}   & {1/2}   \\
 {1/2}   & {1/2}   & 0   & 0   & {1/2}   & {-1/2}
  & 0   & 0   \\
 {-1/2}   & {1/2}   & 0   & 0   & {1/2}   & {1/2}
  & 0   & 0   \\
\end{bmatrix}
\begin{bmatrix}
 {X_A^k }   \\
 {Y_A^k }   \\
 {X_B^k }   \\
 {Y_B^k }   \\
 {X_C^k }   \\
 {Y_C^k }   \\
 {X_D^k }   \\
 {Y_D^k }   \\
\end{bmatrix}
=\begin{bmatrix}
 {X_A^{k+1} }   \\
 {Y_A^{k+1} }   \\
 {X_B^{k+1} }   \\
 {Y_B^{k+1} }   \\
 {X_C^{k+1} }   \\
 {Y_C^{k+1} }   \\
 {X_D^{k+1} }   \\
 {Y_D^{k+1} }   \\
\end{bmatrix}
\end{equation}

\addtocounter{MaxMatrixCols}{10}
\begin{equation}
\label{eq15}
\begin{bmatrix}
 0  & 0  & 0  & {1/2}  & {1/2}  & {-1/2} & 0  & 0  & 0  & {1/2}  & {-1/2}  & {1/2}  \\
 0  & 0  & 0  & {-1/2}  & {1/2}  & {1/2} & 0  & 0  & 0  & {1/2}  & {1/2}  & {-1/2}  \\
 0  & 0  & 0  & {1/2}  & {-1/2}  & {1/2} & 0  & 0  & 0  & {-1/2}  & {1/2}  & {1/2}  \\
 {1/2}  & {-1/2}  & {1/2}  & 0  & 0  & 0 & {1/2}  & {1/2}  & {-1/2}  & 0  & 0  & 0  \\
 {1/2}  & {1/2}  & {-1/2}  & 0  & 0  & 0 & {-1/2}  & {1/2}  & {1/2}  & 0  & 0  & 0  \\
 {-1/2}  & {1/2}  & {1/2}  & 0  & 0  & 0 & {1/2}  & {-1/2}  & {1/2}  & 0  & 0  & 0  \\
 0  & 0  & 0  & {1/2}  & {-1/2}  & {1/2} & 0  & 0  & 0  & {1/2}  & {1/2}  & {-1/2}  \\
 0  & 0  & 0  & {1/2}  & {1/2}  & {-1/2} & 0  & 0  & 0  & {-1/2}  & {1/2}  & {1/2}  \\
 0  & 0  & 0  & {-1/2}  & {1/2}  & {1/2} & 0  & 0  & 0  & {1/2}  & {-1/2}  & {1/2}  \\
 {1/2}  & {1/2}  & {-1/2}  & 0  & 0  & 0 & {1/2}  & {-1/2}  & {1/2}  & 0  & 0  & 0  \\
 {-1/2}  & {1/2}  & {1/2}  & 0  & 0  & 0 & {1/2}  & {1/2}  & {-1/2}  & 0  & 0  & 0  \\
 {1/2}  & {-1/2}  & {1/2}  & 0  & 0  & 0 & {-1/2}  & {1/2}  & {1/2}  & 0  & 0  & 0  \\
\end{bmatrix}
\end{equation}

The left-hand matrix is also the element stiffness matrix
for a planar quadrilateral element, but it is simpler in form to that used in the DM (Eq.\ref{eq7}).
Similarly, the 2D version of stiffness matrix for a planar quadrilateral
element (Eq.\ref{eq14}) can be extended into 3D version (Eq.\ref{eq15}).

\subsubsection{Tri${-}$quad mesh}

The element stiffness matrices used in a tri-quad mesh are those already
given for the triangular and quadrilateral meshes (Eq.\ref{eq10}, \ref{eq11}, \ref{eq14}, \ref{eq15}). Which
of these is used for a particular element in a tri-quad mesh depends only on
that element's type.

%4 The MDM for smoothing planar meshes}
\subsection{The MDM for smoothing planar meshes}

Just as in the DM, the element stiffness matrices can be assembled into a
global matrix, also of size $2n\times 2n$ for a mesh of $n$ nodes. This Jacobi iteration matrix has several forms,
the one shown in Eq.\ref{eq17}
is the most efficient computationally. Therefore it is the one that will
need to be solved iteratively, starting with the original node coordinates
at step 0 and continuing until no node needs to be moved by more than the
given tolerance distance.

\begin{eqnarray}
\label{eq17}
D\cdot
\left \{
\begin{bmatrix}
 {\alpha _{11} }   & {\alpha _{12} }   & {\alpha _{13} }   & \cdots   & {\alpha _{1(2n)} }   \\
 {\alpha _{21} }   & {\alpha _{22} }   & {\alpha _{23} }   & \cdots  & {\alpha _{2(2n)} }   \\
 {\alpha _{31} }   & {\alpha _{32} }   & {\alpha _{33} }   & \cdots     & {\alpha _{3(2n)} }   \\
 \vdots   & \vdots   & \vdots    & \ddots  &  \vdots   \\
 {\alpha _{(2n)1} }   & {\alpha _{(2n)2} }   & {\alpha _{(2n)3} }& \cdots   & {\alpha _{(2n)(2n)} }   \\
\end{bmatrix}
\begin{bmatrix}
 {X_1^k }   \\
 {Y_1^k }   \\
 {X_2^k }   \\
 \vdots   \\
 {Y_n^k }   \\
\end{bmatrix}
\right \}
=\begin{bmatrix}
 {X_1^{k+1} }   \\
 {Y_1^{k+1} }   \\
 {X_2^{k+1} }   \\
 \vdots   \\
 {Y_n^{k+1} }   \\
\end{bmatrix}
\end{eqnarray}

, where$D=diag(1/e{ }_1,\mbox{ }1/e_1 ,\mbox{ }1/e_2 ,\mbox{ }1/e_2 ,\mbox{
}\cdots ,\mbox{ }\cdots ,\mbox{ }1/e_n ,\mbox{ }1/e_n )$

The algorithm for implementing the MDM in 2D has three basic steps:
(1) Search for elements that share a node, for each node;
(2) Assemble of the element stiffness matrices into the iteration matrix;
(3) Solve of the iteration equations until the tolerance distance is reached.

\subsection{The MDM for smoothing surface meshes}

When smooth surface meshes, it is necessary to keep the features of original meshes. Many features preservation approaches have been proposed\cite{7., 15., 17.}. A popular methods is to classify all the nodes of
a mesh into four types: boundary, corner, ridges and smooth nodes; and then boundary nodes and corner nodes are fixed while smooth node can be adjusted on the
whole mesh and ridge nodes can only be relocated along the ridges.

In order to preserve features, we firstly detect the corner nodes and ridge nodes via Jiao's approach \cite{15.},
and fix them as constrained nodes together with the boundary nodes although
the ridge nodes can be moved along ridges; and then after obtaining the
relocated positions based on the Jacobi iteration matrix, we project the
these new nodes onto the original mesh to get the mapped ones which are
still the candidates of the resulting nodes in this iteration; and thirdly,
we check whether there exists inverted elements in the incident faces
of the mapped node. If does, recover it; otherwise, update the target node with the mapped position(Algorithm.\ref{alg:1:MDM3D}).

Since the MDM is an iterative smoothing algorithm, we have to conduct some
indicators to judge when the meshes are smoothed enough and then end the
iteration. We adopt two mesh quality indicators, the mean quality of mesh (MQ) and mean square error (MSE) of all elements, to show target
meshes are smoothed enough and iterations can break. This can be presented
as:
\[
\left\{\begin{array}{r@{,\quad}l}
{if}\left\{ \begin{array}{c}
 {\mbox{MQ}^{k+1}-\mbox{MQ}^k}<\varepsilon _{mq} \\
 {\mbox{MSE}^{k+1}-\mbox{MSE}^k}<\varepsilon _{mse}
\end{array}\right.& \mbox{stop} \\
 {otherwise} & \mbox{continue to iterate}
\end{array}\right.
\]
, where $\varepsilon _{mq} $ and $\varepsilon _{mse} $ are two user-specified thresholds. Noticeably, the above criteria only works when there are no inverted elements.
\begin{algorithm}[H]
\renewcommand{\algorithmicrequire}{\textit{Input:}}
\renewcommand\algorithmicensure {\textit{Output:}}
\caption{The MDM for smoothing surface meshes}
\label{alg:1:MDM3D}
\begin{algorithmic}[1]

\STATE Search all incident faces $F(v_i )$ for each node $v_i $;
\STATE Assemble the $3n\times 3n$ Jacobi iteration matrix ${\rm {\bf B}}=(b_{ij} )_{0\le i<3n,0\le j<3n} $;
\WHILE{iteration not converge}
    \STATE Estimate or update normal at each node $v_i $
    \FOR{each node $v_i $}
        \IF{$v_i $ is not constrained}
            \STATE Calculate the relocated node $v_i^{new} $ based on ${\rm {\bf B}}$:
            \STATE Project $v_i^{new} $onto $F(v_i )$ to obtain the mapped node $v_i^{map} $;
            \STATE Check inverted element in $F(v_i^{map} )$. If exists, recover $\mbox{ }v_i^{map} \leftarrow v_i $
        \ENDIF
    \ENDFOR
\STATE Update all nodes: $v_i \mbox{ }\leftarrow v_i^{map} $
\ENDWHILE

\end{algorithmic}
\end{algorithm}

\subsubsection{Normal of vertices}
\label{sec:5.1}

The direction of each vertex is closely related to its incident faces, for
that the normal of each vertex is nearly vertical to the normal of any face
of its incident elements. Therefore, the unit normal of each faces in a mesh
should be calculated firstly by computing the planar equation of every face.
Suppose there are $m$ faces shares the vertex $v$, and the normal of $v$ can be
obtained by solving the following $m\times 3$ linear equations
{\rm {\bf N}}\mbox{x}={\rm {\bf 1}},
 where ${\rm {\bf N}}$ is a $m\times 3$ matrix whose $i^{th}$ row is the
unit normal of the $i^{th}$ incident face of the vertex $v$, and ${\rm {\bf
1}}=(1,1,\mbox{...,1})$ is a vector of length $m$. Since ${\rm {\bf N}}$ may be
over- or under-determined, the solution is in least squares sense and can be
solved by the singular value decomposition (SVD).

\subsubsection{Identifying points}
\label{sec:5.2}

We adopt Jiao's algorithm \cite{15.} to detect features based on eigenvalues
analysis of a symmetric positive semi-definite matrix ${\rm {\bf A}}$:
${\rm {\bf A}}={\rm {\bf N}}^T{\rm {\bf WN}}$,
where ${\rm {\bf N}}$ is a $m\times 3$ matrix we denote in above section,
and ${\rm {\bf W}}$ be a $m\times m$ diagonal matrix with $W_{ii} $ equal to
the weight. If users do not consider the weights, ${\rm {\bf W}}$ can be ignored and ${\rm {\bf A}}$ is then simplified
as ${\rm {\bf N}}^T{\rm {\bf N}}$. Let $\lambda _1 $, $\lambda _2 $ and
$\lambda _3 $ ($\lambda _1 \ge \lambda _2 \ge \lambda _3 )$ be the three
eigenvalues of ${\rm {\bf A}}$.
The relative sizes of the eigenvalues $\lambda _i $ of ${\rm {\bf A}}$ are
closely related to the local flatness at a vertex. In general, ${\rm {\bf
A}}$ has three large eigenvalues at a corner, two large ones at a ridge, and
one large one at a smooth point. Hence, the corners and ridges can be
recognized by comparing $\lambda _3 /\lambda _1 $ and $\lambda _2 /\lambda
_1 $ against some thresholds: 
$\mbox{if }\lambda _3 /\lambda _1 \ge \chi _c ,$  $v\mbox{ is at corner};$
$\mbox{if }\lambda _2 /\lambda _1 \ge \chi _r ,$  $v\mbox{ is on a ridge}$
, where $\chi _c $ and $\chi _r $ are two given thresholds.

\subsubsection{Projecting }
\label{sec:5.3}

After relocating position in an
iteration step, smooth node will be projected onto its incident faces of the
original mesh alone the updated normal. Since that, we just project the relocated
point onto its last incident faces (neighborhood), there maybe not exists a
mapped point on the original mesh inside the neighborhood of the above node.
Thus, let $k$ be the number of mapped points on
the incident faces, then we have:
(1) if $k=0$, ignore the relocated position and recover its last coordinates;
(2) if $k=1$, adopt the only mapped point as the new position;
(3) if $k>1$, select the nearest mapped point to the relocated position as
the new position.

% 6 Test applications of the MDM
\section{Test applications of the MDM}
\label{sec:6}

\subsection{Mesh quality}
\label{sec:6.1}

The simplest way to measure
mesh quality is to calculate distortion values for each of the mesh elements.
The distortion value for a triangular element should measure how close that
triangle is to equilateral. One appropriate measure, \textit{$\alpha $}, was proposed by Lee
and Lo\cite{20.}. For the triangle ABC shown in Fig.\ref{fig:1:DM:T}:
\[
\alpha =2\sqrt 3 \frac{\left\| {\mbox{CA}\times \mbox{CB}} \right\|}{\left\|
{\mbox{CA}} \right\|^2+\left\| {\mbox{AB}} \right\|^2+\left\| {\mbox{BC}}
\right\|^2}
\]
The value of \textit{$\alpha $} lies between 0 and 1; \textit{$\alpha $} = 0 when A, B and C are collinear;
\textit{$\alpha $} = 1 when ABC is equilateral.

The distortion value for a quadrilateral element should measure how close
that quadrilateral is to square. In this paper we use the measure \textit{$\lambda $} proposed
by Hua \cite{14.}; this applies only to convex quadrilaterals. For the
quadrilateral ABCD shown in Fig.\ref{fig:3:DM:Q}:
\[
\lambda =2\sqrt[4]{\frac{\left\| {\mbox{AB}\times \mbox{AD}} \right\|\cdot
\left\| {\mbox{BC}\times \mbox{BA}} \right\|\cdot \left\| {\mbox{CD}\times
\mbox{CB}} \right\|\cdot \left\| {\mbox{DA}\times \mbox{DC}}
\right\|}{(\left\| {\mbox{AB}} \right\|^2+\left\| {\mbox{AD}}
\right\|^2)(\left\| {\mbox{BC}} \right\|^2+\left\| {\mbox{BA}}
\right\|^2)(\left\| {\mbox{CD}} \right\|^2+\left\| {\mbox{CB}}
\right\|^2)(\left\| {\mbox{DA}} \right\|^2+\left\| {\mbox{DC}}
\right\|^2)}}
\]
The value of \textit{$\lambda $} lies between 0 and 1; \textit{$\lambda $} = 0 when any three nodes are collinear,
i.e., when ABCD is in fact a triangle; \textit{$\lambda $} = 1 when ABCD is square.

\subsection{The test applications}
\label{sec:6.2}

\begin{figure}[H]
\centering
\subfigure[Original tri mesh]{
  \label{fig:4a}       % Give a unique label
  \includegraphics[width=37mm]{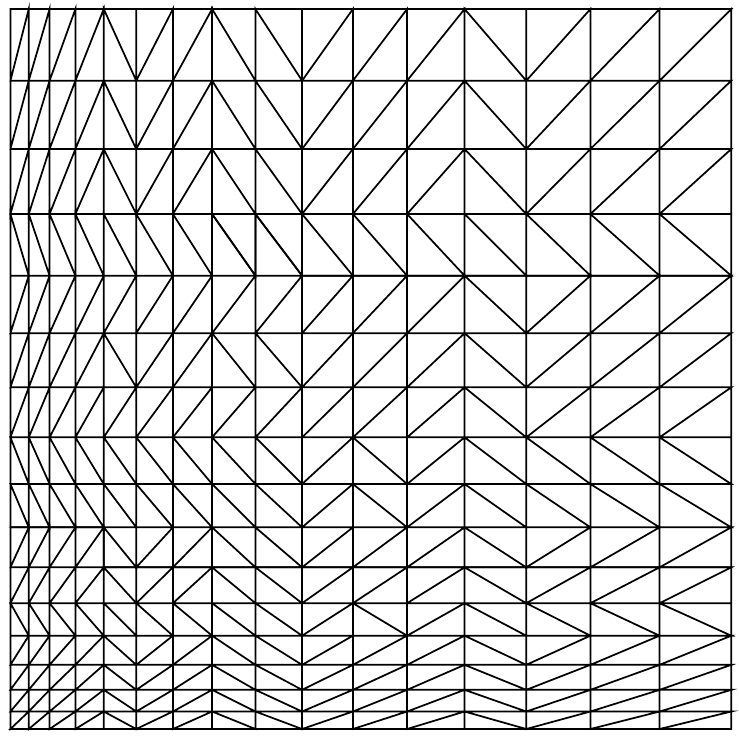} }
\hspace{0cm}
\subfigure[Smoothed by LS/MDM]{
  \label{fig:4b}       % Give a unique label
  \includegraphics[width=37mm]{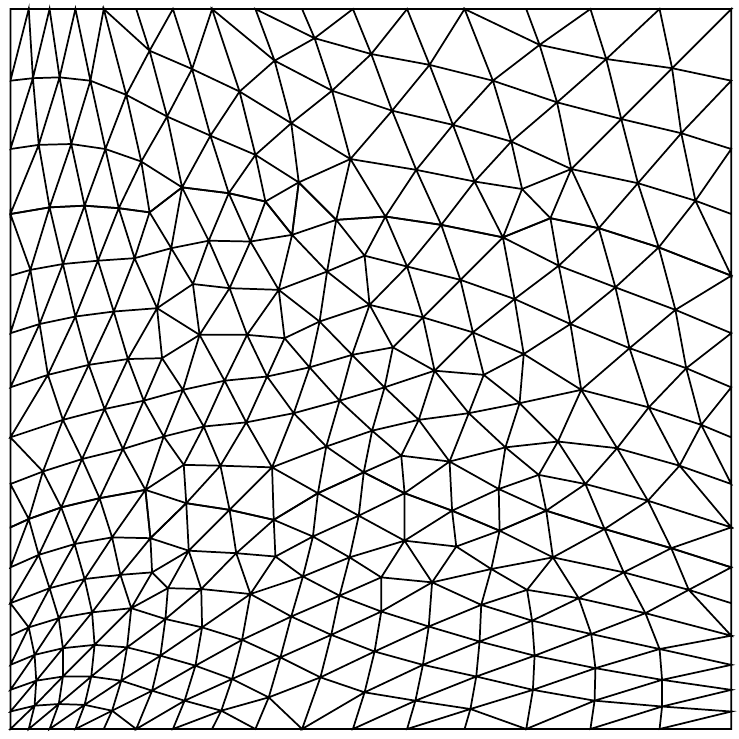} }
\hspace{0cm}
\caption{Smoothing for planar triangular mesh}
\label{fig:4:Smooth:T}       % Give a unique label
\end{figure}

\begin{figure}[H]
\centering
\subfigure[Original quad mesh]{
  \label{fig:5a}       % Give a unique label
  \includegraphics[width=37mm]{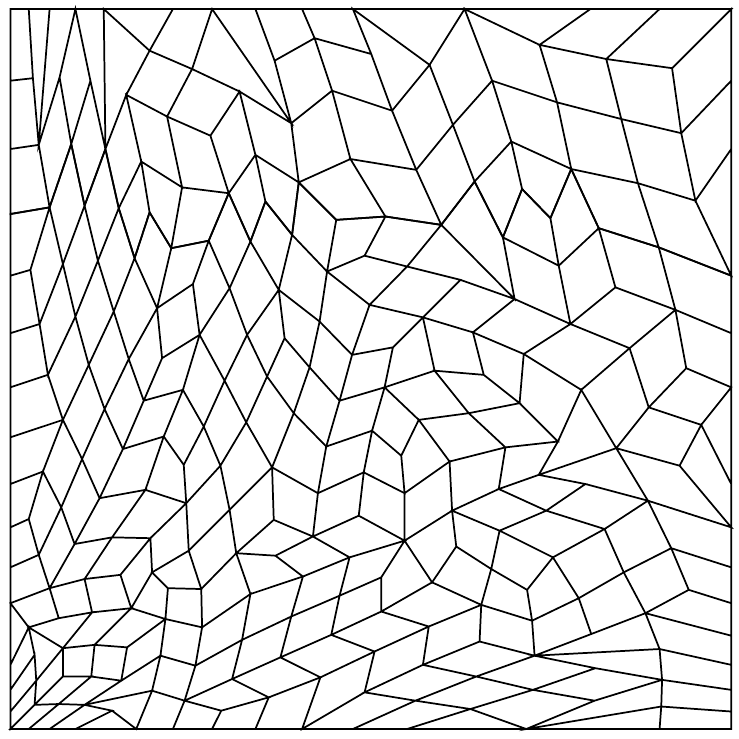} }
\hspace{0cm}
\subfigure[Smoothed by LS/MDM]{
  \label{fig:5b}       % Give a unique label
  \includegraphics[width=37mm]{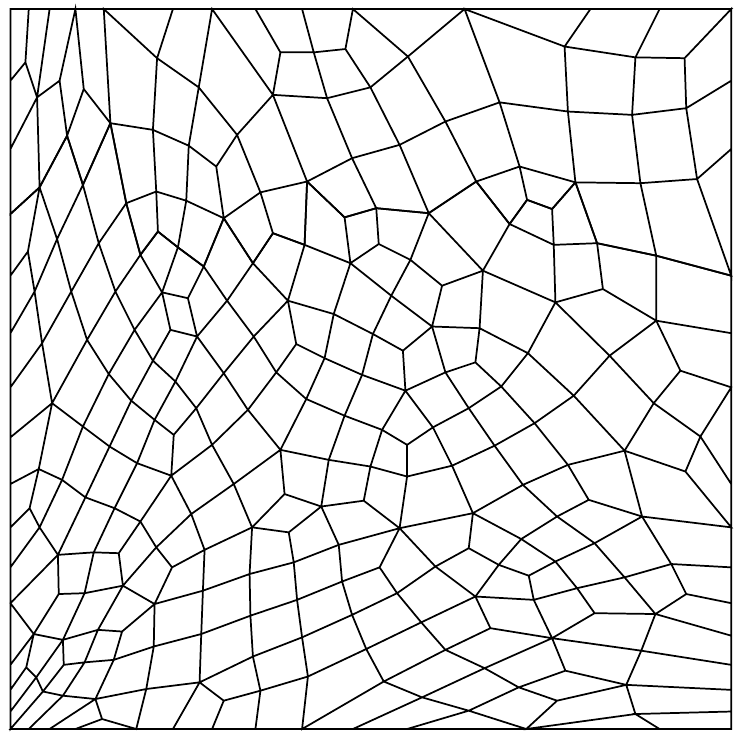} }
\hspace{0cm}
\caption{Smoothing for planar quadrilateral mesh}
\label{fig:5:Smooth:Q}       % Give a unique label
\end{figure}

\begin{figure}[H]
\centering
\subfigure[Tri-dominant mesh]{
  \label{fig:6a}       % Give a unique label
  \includegraphics[width=35mm]{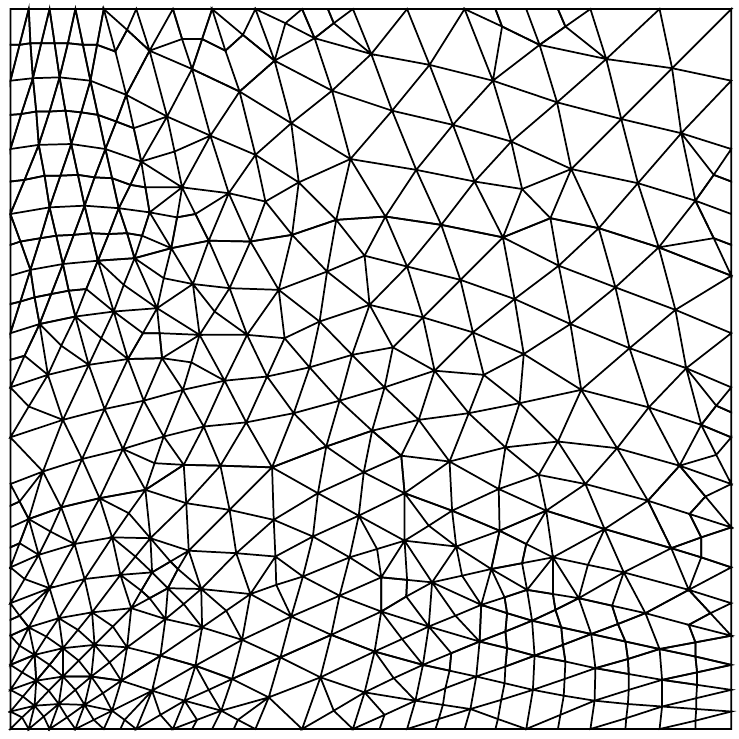} }
\hspace{0cm}
\subfigure[Smoothed by LS]{
  \label{fig:6b}       % Give a unique label
  \includegraphics[width=35mm]{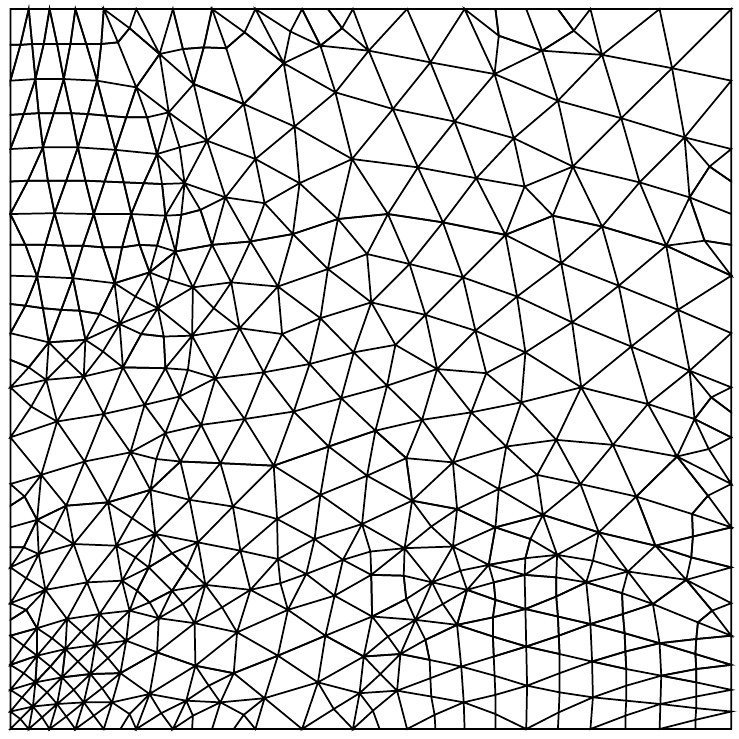} }
\hspace{0cm}
\subfigure[Smoothed by MDM]{
  \label{fig:6c}       % Give a unique label
  \includegraphics[width=35mm]{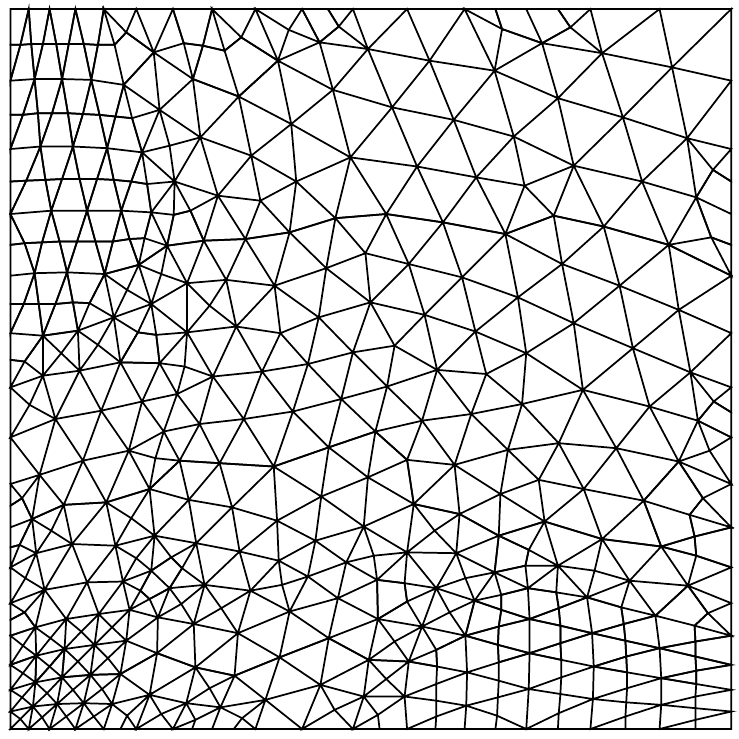} }
\hspace{0cm}
\subfigure[Quad-dominant mesh]{
  \label{fig:7a}       % Give a unique label
  \includegraphics[width=35mm]{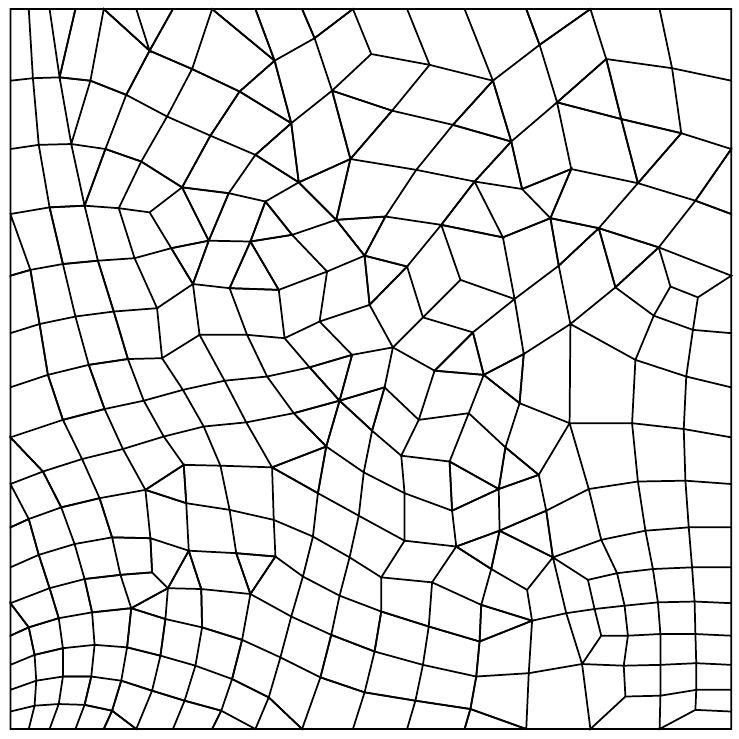} }
\hspace{0cm}
\subfigure[Smoothed by LS]{
  \label{fig:7b}       % Give a unique label
  \includegraphics[width=35mm]{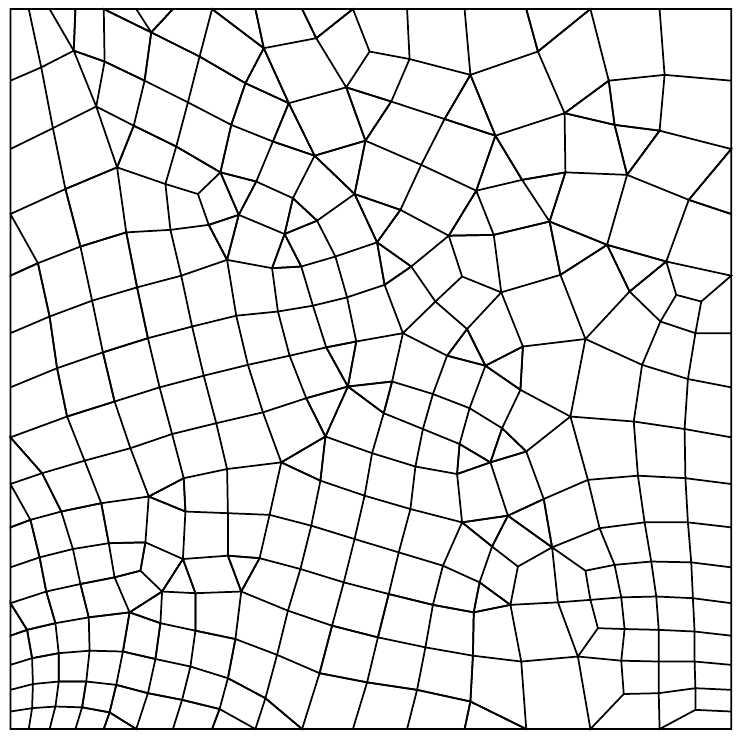} }
\hspace{0cm}
\subfigure[Smoothed by MDM]{
  \label{fig:7c}       % Give a unique label
  \includegraphics[width=35mm]{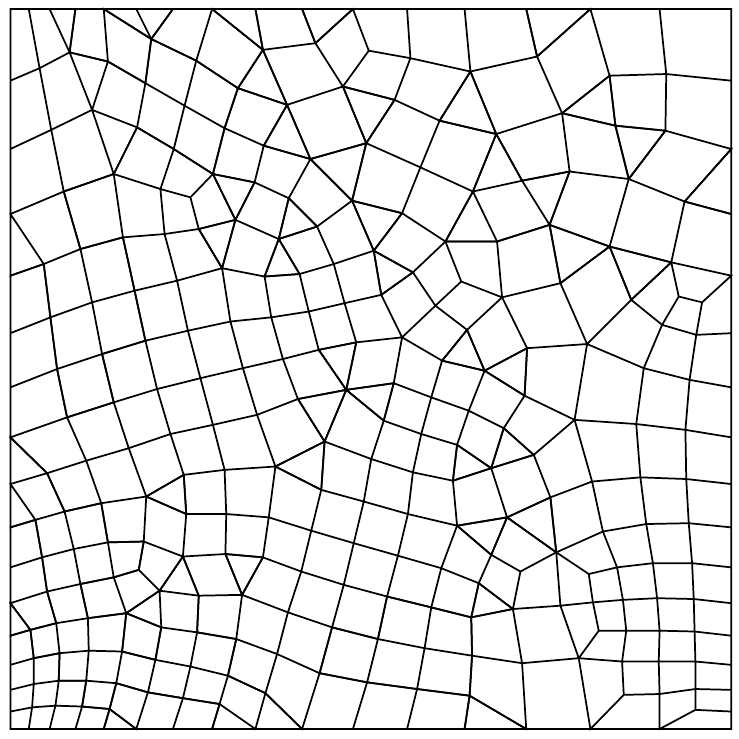} }
\caption{Smoothing for planar tri-quad meshes}
\label{fig:7:Smooth:T_Q}       % Give a unique label
\end{figure}

For smoothing planar meshes, a number of test meshes were created and
smoothed (Figs. \ref{fig:4:Smooth:T}, \ref{fig:5:Smooth:Q}, \ref{fig:7:Smooth:T_Q}). The mesh quality results for tri-quad meshes before and after smoothing are given in Tables \ref{tab:1}.
For smoothing surface meshes, a tri and a quad surface mesh are created by planar
triangulations\cite{19., 20., 25.} and then interpolated to surface (Fig.\ref{fig:8:Smooth:Surface}).

\begin{table}[H]
\centering
\caption{Smoothing for planar tri-quad meshes(T: tri; Q: quad)}
\label{tab:1}       % Give a unique label
\begin{tabular}{lllll|llll}
\hline\noalign{\smallskip}
  &  \multicolumn{4}{c}{T-dominant} &  \multicolumn{4}{c}{Q-dominant}  \\
 \cline{2-9}
& T-MQ &  T-MSE &  Q-MQ & Q-MSE & T-MQ &  T-MSE &  Q-MQ & Q-MSE \\
\noalign{\smallskip}\hline\noalign{\smallskip}
Original &
0.8688 &
0.1188 &
0.8211 &
0.0713 &
0.9079 &
0.1243 &
0.9194 &
0.0739  \\

LS&
0.8756 &
0.1083 &
0.8464 &
0.0571 &
0.9064 &
0.0936 &
0.9554 &
0.0568  \\

MDM&
0.8794 &
0.1083 &
0.8306 &
0.0583 &
0.9394 &
0.0737 &
0.9576 &
0.0568  \\
\noalign{\smallskip}\hline
\end{tabular}
\end{table}

A tri-dominant mixed  mesh is generated by dividing each sliver triangle and its neighbors in the original triangular mesh into a smaller triangle and a quadrilateral (Fig.\ref{fig:8e}). Another quad-dominant mixed mesh is created by pairing triangles in 2D and then interpolating (Fig.\ref{fig:8g}).

The smoothed surface meshes are shown in Fig.\ref{fig:8:Smooth:Surface}. The user-specified
thresholds $\varepsilon _{mq} $ and $\varepsilon _{mse} $
are set as 10$^{-6}$, 10$^{-4}$, 10$^{-5}$ and 10$^{-4}$ for triangular, quadrilateral, tri-dominant
and quad-dominant meshes, respectively. And correspondingly, the MDM converges at
the steps 43, 56, 33 and 13. The mesh quality results before and after
smoothing are given in Table \ref{tab:2}.

\subsection{Convergence analysis}
\label{sec:6.3}

In this section, we analyze the convergence of the MDM. When smooth the
planar meshes, test examples show that the MDM does converge. We have not
given the mathematical proof for it in theory. For smoothing surface meshes,
the MDM is much more complicated. As mentioned above, the mean quality of mesh (MQ) and mean square error (MSE) of
element qualities are used to indicate the quality of whole mesh, we can
calculate and compare the two indicators in increasing smoothing iterations to analyze the converge.

\begin{figure}[H]
\centering
\subfigure[MQs of smoothing tri mesh]{
  \label{fig:9a}       % Give a unique label
  \includegraphics[width=5.5cm]{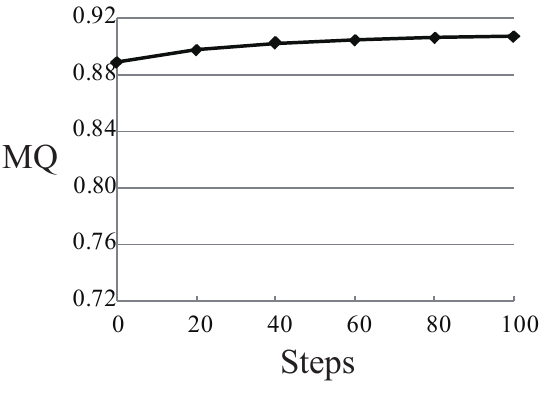} }
\hspace{0cm}
\subfigure[MSEs of smoothing tri mesh]{
  \label{fig:9b}       % Give a unique label
  \includegraphics[width=5.5cm]{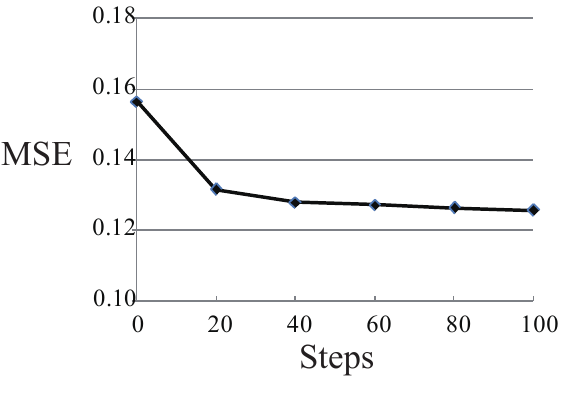} }
\hspace{0cm}

\subfigure[MQs of smoothing quad mesh]{
  \label{fig:10a}       % Give a unique label
  \includegraphics[width=5.5cm]{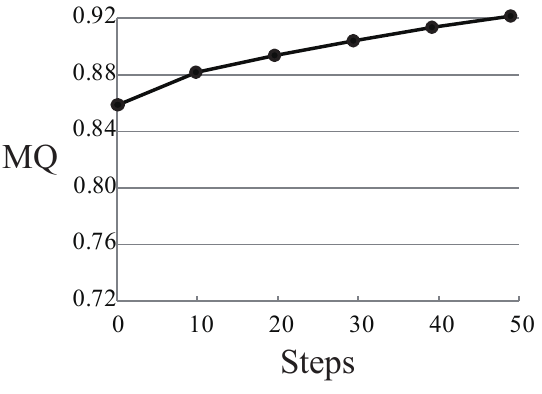} }
\hspace{0cm}
\subfigure[MSEs of smoothing quad mesh]{
  \label{fig:10b}       % Give a unique label
  \includegraphics[width=5.6cm]{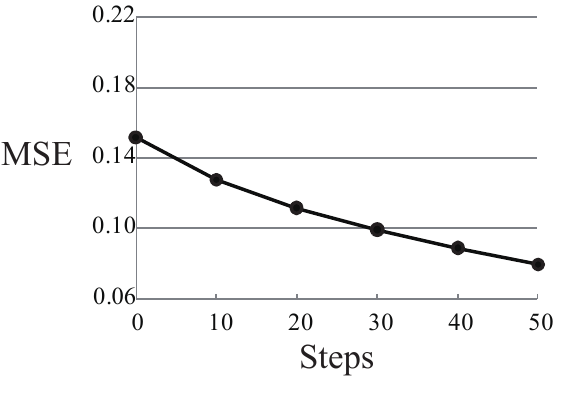} }
\hspace{0cm}

\subfigure[MQs of smoothing tri-domi. mesh]{
  \label{fig:11a}       % Give a unique label
  \includegraphics[width=5.5cm]{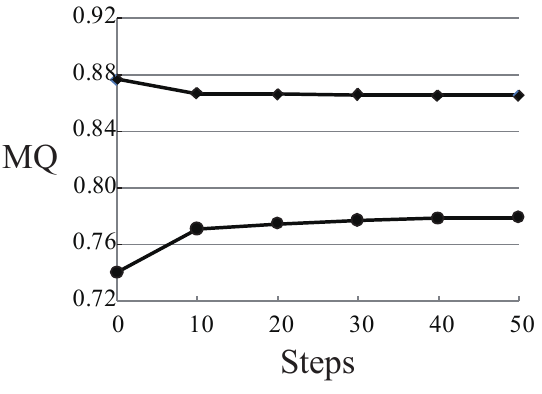} }
\hspace{0cm}
\subfigure[MSEs of smoothing tri-domi. mesh]{
  \label{fig:11b}       % Give a unique label
  \includegraphics[width=5.6cm]{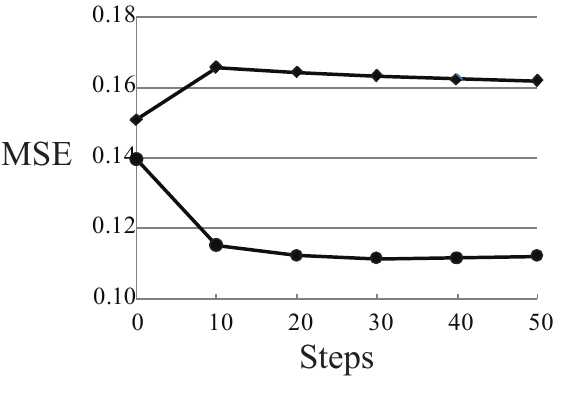} }
\hspace{0cm}
\caption{Mean quality(MQ) and mean square error of quality(MSE) when smooth surface triangular, quadrilateral and tri-domi. meshes. $\blacklozenge$:triangles, $\bullet$:quadrilterals}
\label{fig:11:Convergence:T_Q}       % Give a unique label
\end{figure}

\begin{figure}[H]
\centering
\subfigure[Original triangular mesh]{
  \label{fig:8a}       % Give a unique label
  \includegraphics[width=55mm]{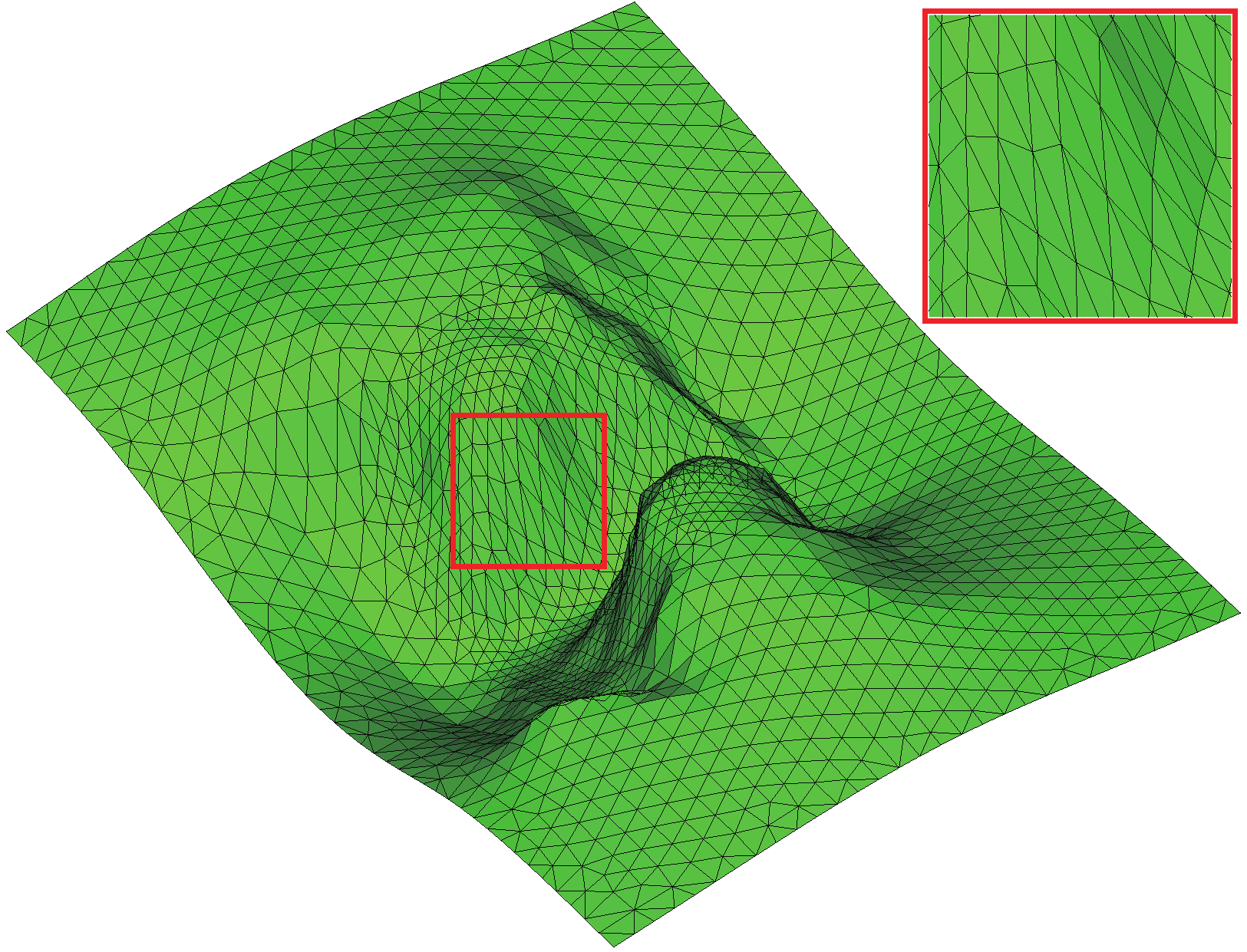} }
\hspace{0cm}
\subfigure[Smoothed by MDM]{
  \label{fig:8b}       % Give a unique label
  \includegraphics[width=55mm]{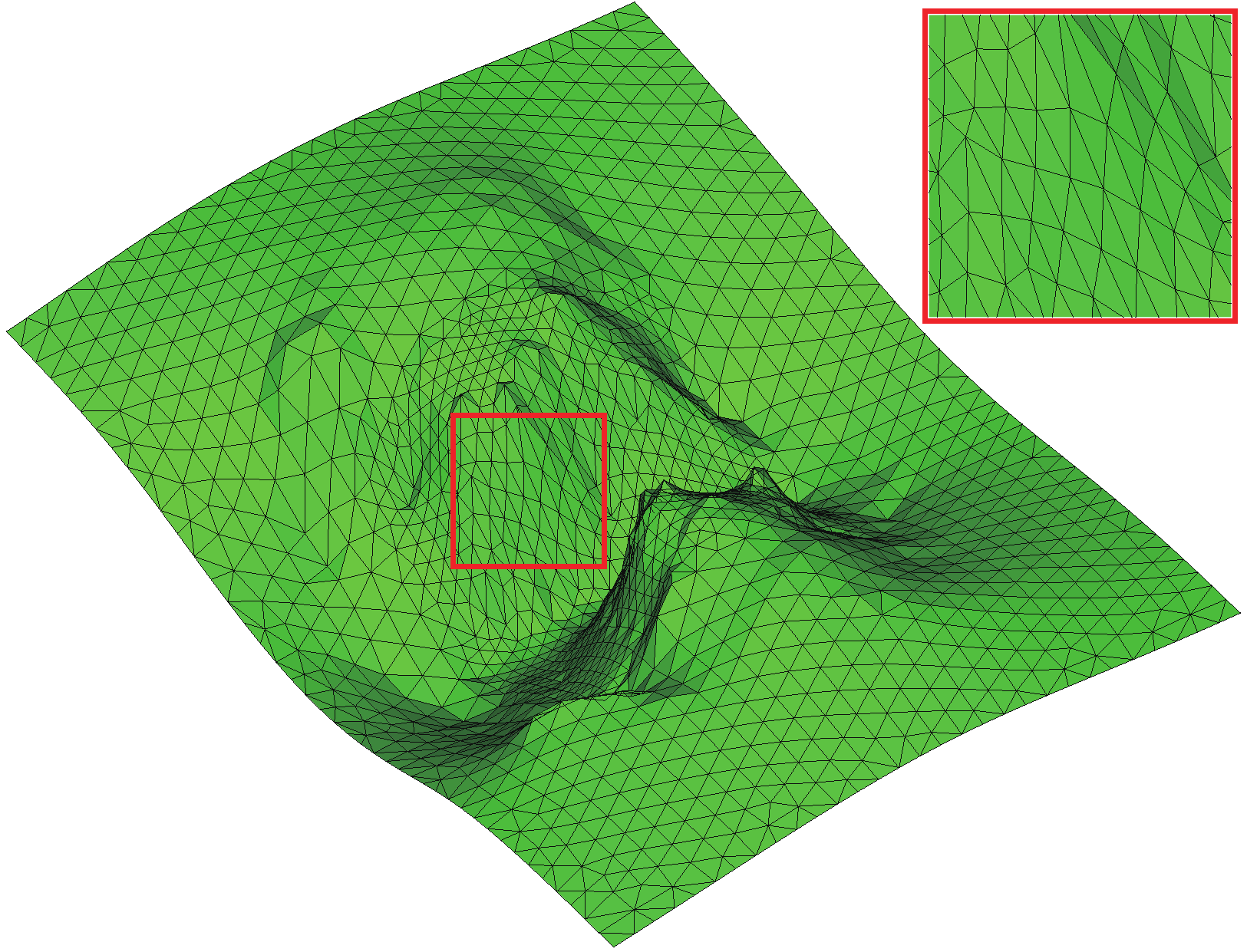} }
\hspace{0cm}

\subfigure[Original quadrilateral mesh]{
  \label{fig:8c}       % Give a unique label
  \includegraphics[width=55mm]{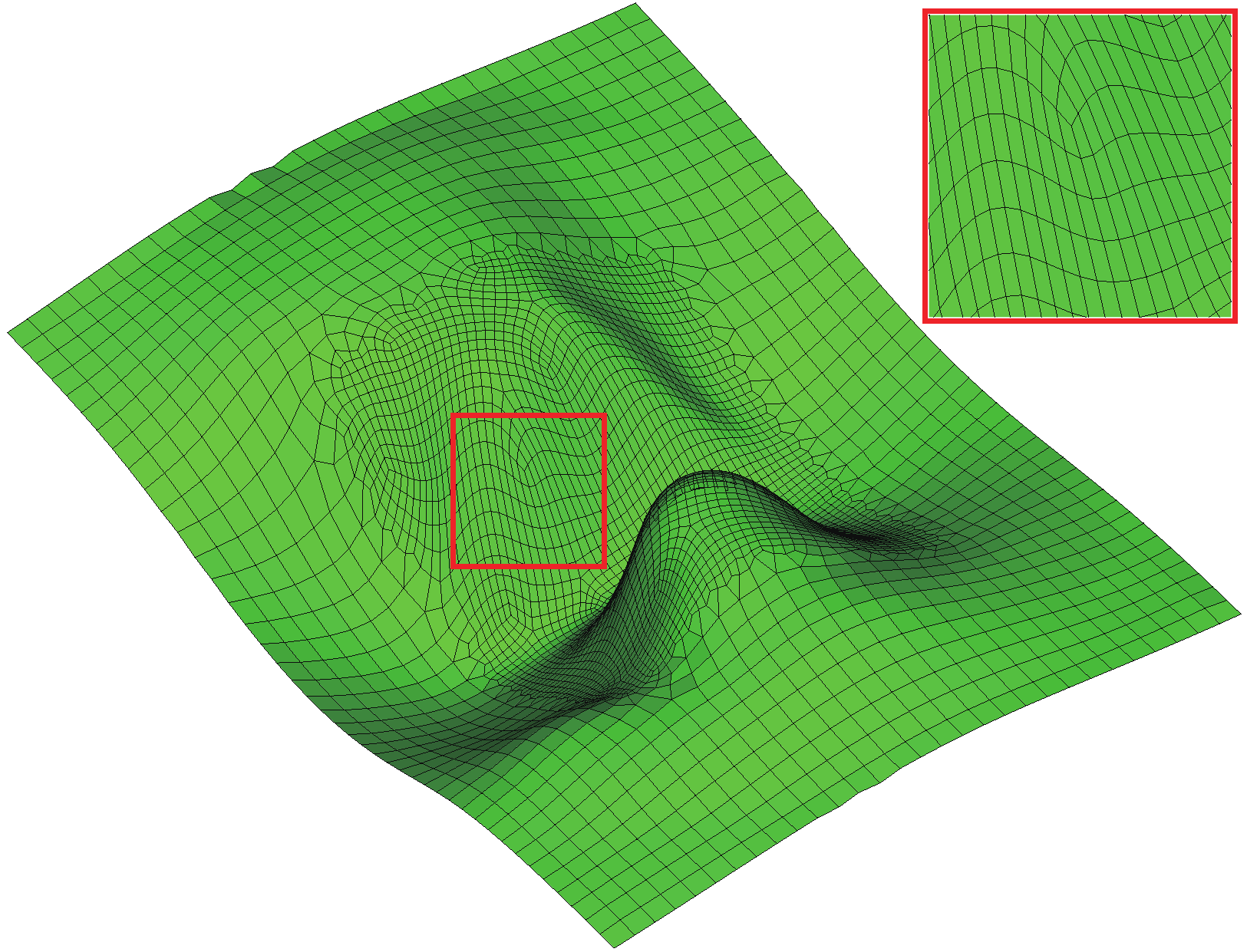} }
\hspace{0cm}
\subfigure[Smoothed by MDM]{
  \label{fig:8d}       % Give a unique label
  \includegraphics[width=55mm]{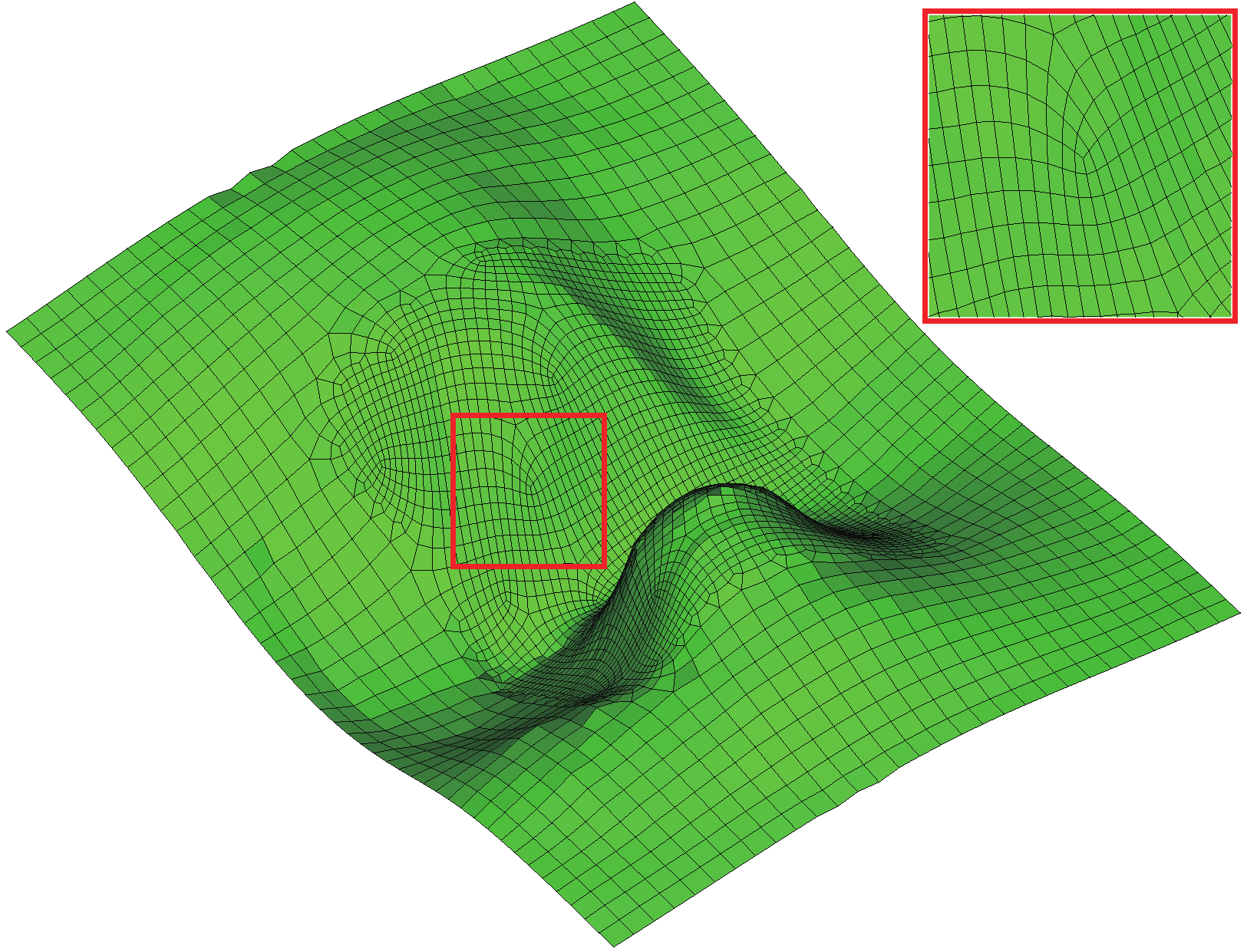} }
\hspace{0cm}

\subfigure[Original tri-dominant mesh]{
  \label{fig:8e}       % Give a unique label
  \includegraphics[width=55mm]{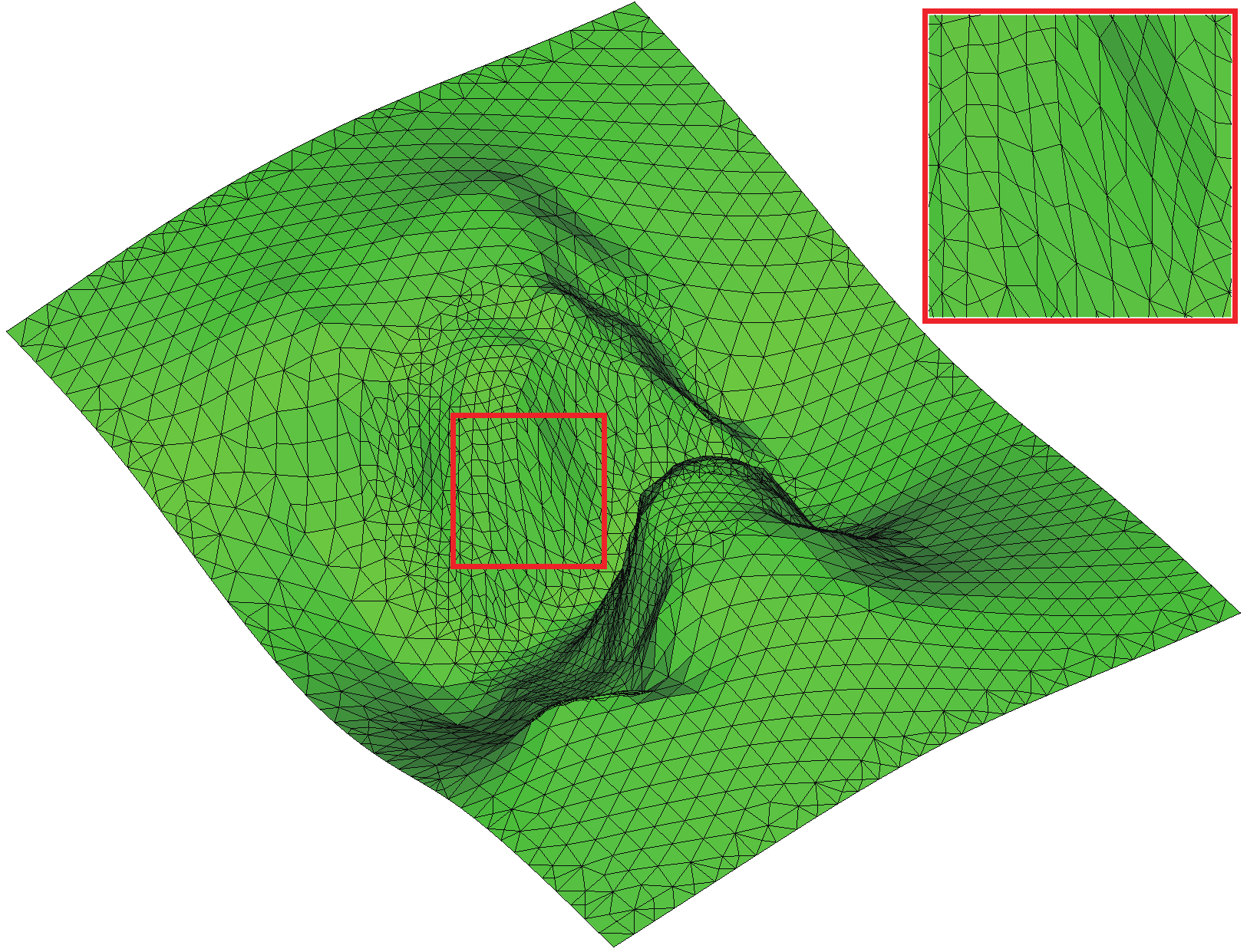} }
\hspace{0cm}
\subfigure[Smoothed by MDM]{
  \label{fig:8f}       % Give a unique label
  \includegraphics[width=55mm]{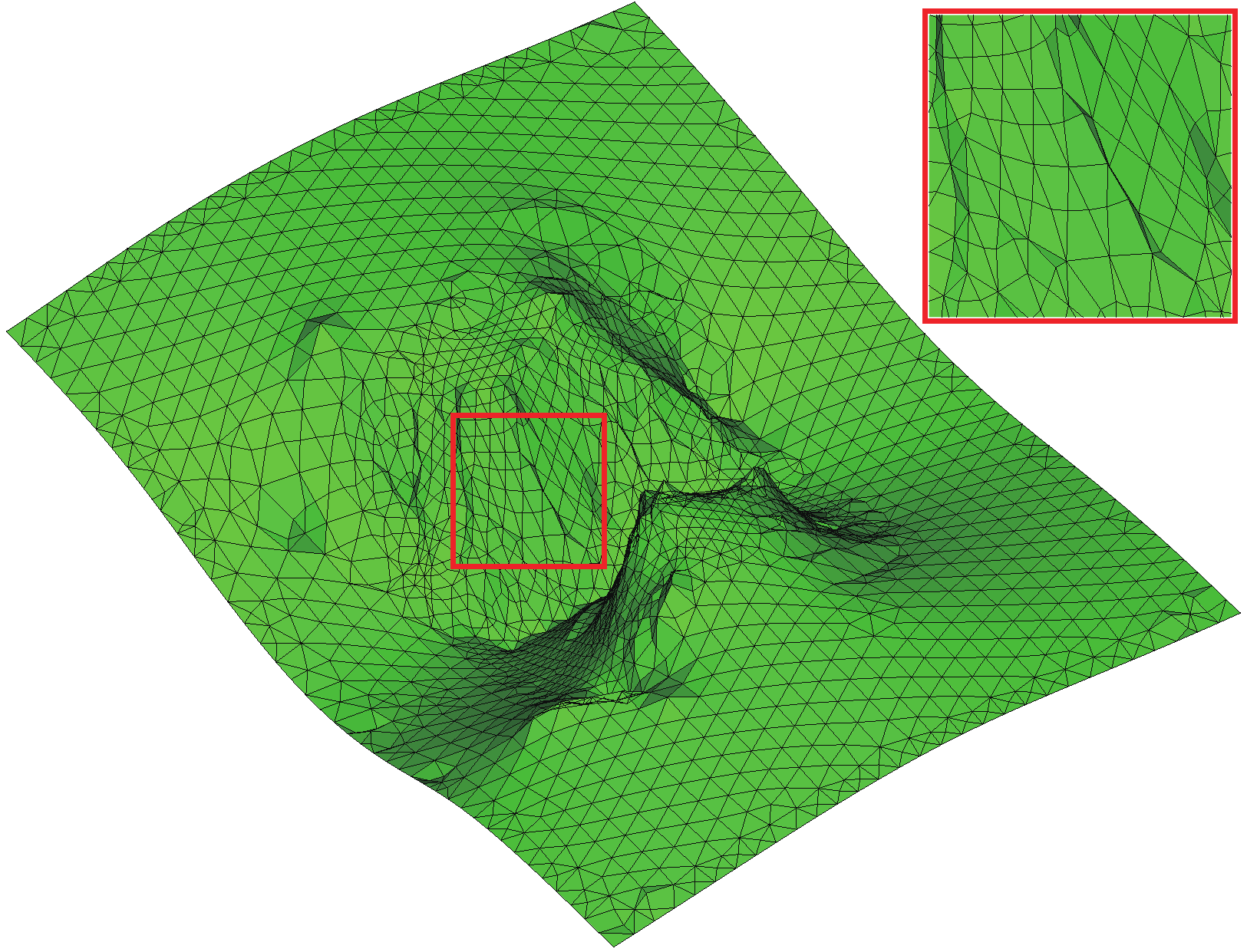} }
\hspace{0cm}

\subfigure[Original quad-dominant mesh]{
  \label{fig:8g}       % Give a unique label
  \includegraphics[width=55mm]{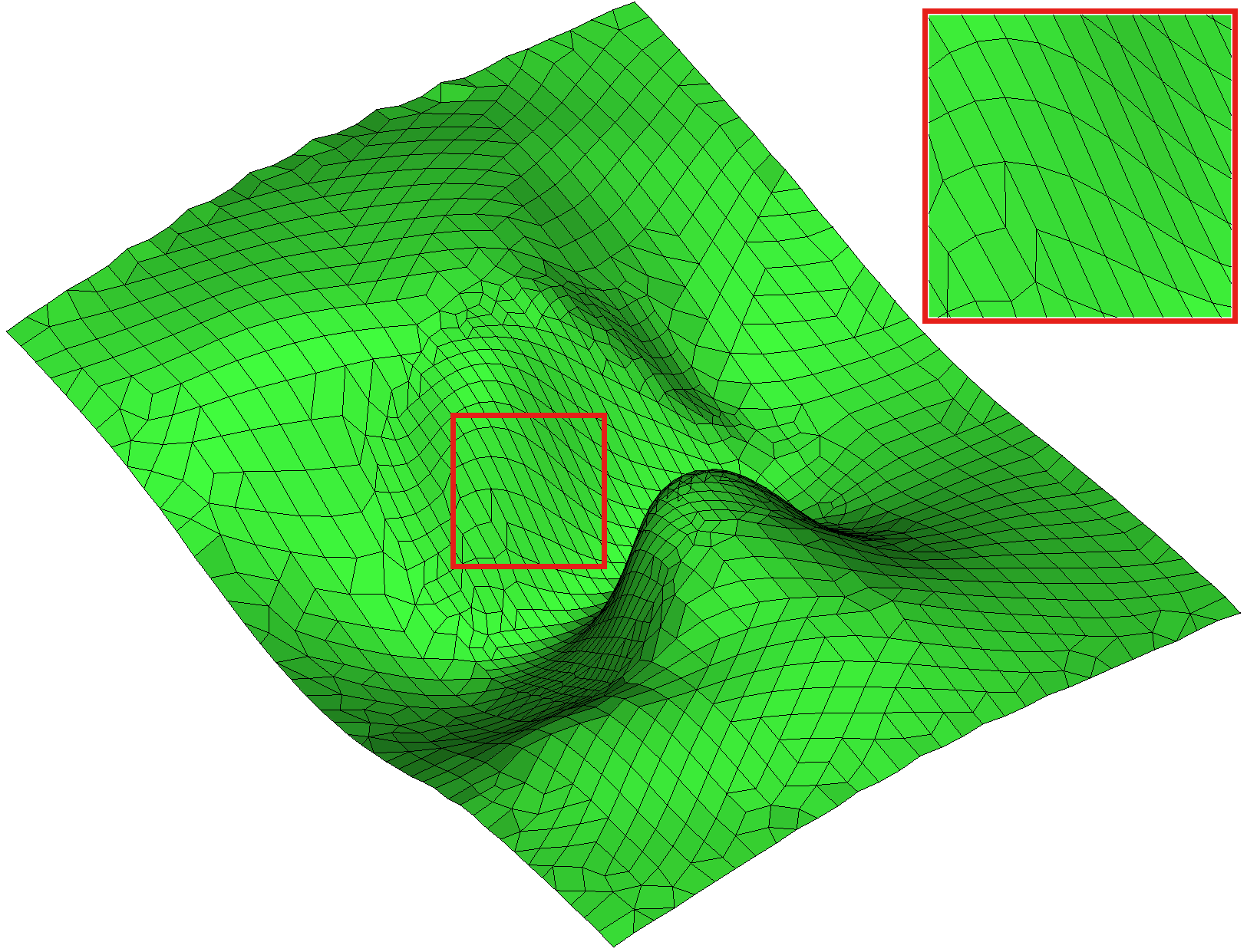} }
\hspace{0cm}
\subfigure[Smoothed by MDM]{
  \label{fig:8h}       % Give a unique label
  \includegraphics[width=55mm]{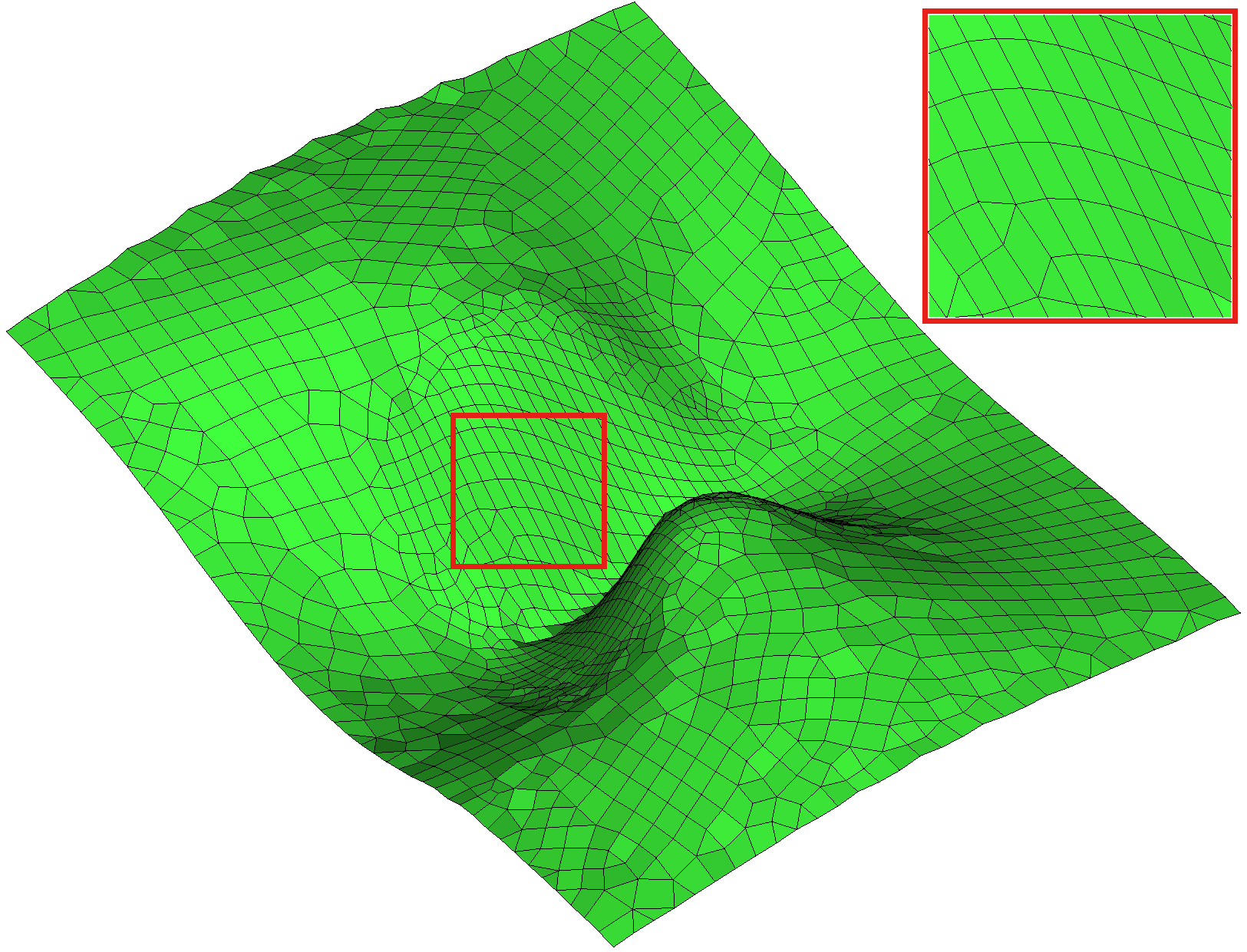} }
\hspace{0cm}
\caption{Smoothing surface meshes by MDM}
\label{fig:8:Smooth:Surface}       % Give a unique label
\end{figure}

The triangular surface (Fig.\ref{fig:8a}) is smoothed by MDM in 20, 40, 60, 80 and 100
iterations. The MQ and MSE of smoothed meshes in different
iteration stages are listed in Table \ref{tab:2}. And corresponding scatter diagrams
are drawn in Fig. \ref{fig:9a} and \ref{fig:9b}. It is clear that the mean quality of mesh
ascends while MSE descends during increasing iterations. The quadrilateral
mesh is smoothed by MDM in 10, 20, 30, 40 and 50 iterations. From Table \ref{tab:2} and
Fig. \ref{fig:10a} and \ref{fig:10b}, we can also receive the same conclusions as that of triangular
mesh. Similarly, the tri-dominant mixed mesh is smoothed in 10, 20, 30, 40 and 50 iterations(Table \ref{tab:2}, Fig. \ref{fig:11a} and \ref{fig:11b}), and the quad-dominant mixed mesh is smoothed in 5, 10, 15, 20 and 25 iterations(Table \ref{tab:2}).

According to the scatter diagrams of MQ and MSE,
we can learn that: for both triangular and quadrilateral surface meshes, the
MQ always increases and the MSE decreases during
smoothing. But the magnitude and rate of change are becoming smaller and
smaller. Thus, we may in theory receive the conclusion that the smoothing
iterations will converge after some steps. This can be also concluded for quad-dominant mixed mesh.

For the tri-dominant mixed mesh, the quality of triangles always decreases while that of
quads increases. Noticeably, in beginning iteration steps, both of the above
qualities change dramatically. Different from the mean qualities, the MSE is
much more complex: the MSE of triangles significantly ascends in beginning
steps, and then descends stably; the magnitude and rate of change are
becoming smaller and smaller. The MSE of quadrilaterals seems to always
decline, but obviously, there is an inflection point at steps 30(the MSEs at
both steps 20 and steps 40 are bigger than that at steps 30); hence, there
must be a convergence point around steps 30. 
Our tests (Table \ref{tab:2}) proves the above conclusion: the
MDM converges at steps 33 when the thresholds $\varepsilon _{mq} $ and
$\varepsilon _{mse} $ are set as 10$^{-5}$. In Table \ref{tab:2}, each stage includes 5, 10 or 20 steps.

\begin{table}[H]
\centering
\caption{Qualities for smoothing surface meshes in increasing iteration steps}
\label{tab:2}       % Give a unique label
\begin{tabular}{lllllllll}
\hline\noalign{\smallskip}
Meshes &
&
Indicator&
Stage 0&
Stage 1&
Stage 2&
Stage 3&
Stage 4&
Stage 5 \\

\noalign{\smallskip}\hline\noalign{\smallskip}

\multirow{2}*{Tri mesh}
&
&
MQ&
0.8891&
0.8980&
0.9026&
0.9049&
0.9067&
0.9073 \\

&
&
MSE&
0.1565&
0.1317&
0.1281&
0.1272&
0.1264&
0.1257 \\

\multirow{2}*{Quad mesh}
&
&
MQ&
0.8592&
0.8814&
0.8929&
0.9027&
0.9122&
0.9198 \\

&
&
MSE&
0.1466 &
0.1237&
0.1088&
0.0973&
0.0874&
0.0785 \\

\multirow{2}*{Tri of tri-domi.}
&
&
MQ&
0.8769&
0.8672&
0.8663&
0.8660&
0.8659&
0.8658 \\

&
&
MSE&
0.1509&
0.1657&
0.1644&
0.1633&
0.1625&
0.1619 \\

\multirow{2}*{Quad of tri-domi.}
&
&
MQ&
0.7406&
0.7711&
0.7750&
0.7772&
0.7785&
0.7792 \\

&
&
MSE&
0.1397&
0.1153&
0.1123&
0.1115&
0.1116&
0.1122 \\

\multirow{2}*{Tri of quad-domi.}
&
&
MQ&
0.9187&
0.9417&
0.9435 &
0.9449 &
0.9466&
0.9481 \\

&
&
MSE&
0.1066&
0.0884&
0.0839&
0.0813 &
0.0777 &
0.0747 \\

\multirow{2}*{Quad of quad-domi.}
&
&
MQ&
0.8220&
0.8625 &
0.8721&
0.8788&
0.8842&
0.8887  \\

&
&
MSE&
0.1432 &
0.1323&
0.1235&
0.1162&
0.1102&
0.1048 \\
\noalign{\smallskip}\hline
\end{tabular}
\end{table}

\subsection{Tests assessment and summary}
\label{sec:6.4}
1) For the two topologically uniform planar meshes, i.e., the meshes with the
same type of element throughout, there is effectively no difference between
Laplacian smoothing and MDM. (This was referred to earlier, in
the description of the original DM.) Both these methods give smoothed meshes
that are markedly better than the original unsmoothed meshes.

2) For the triangular elements in the planar tri-dominant mixed mesh, the MDM
outperforms Laplacian smoothing while for the quadrilateral elements,
the reverse is true (Table \ref{tab:1}). However, for planar quad-dominant mixed mesh, the MDM perfectly outperforms Laplacian smoothing for both triangular and quadrilateral elements.

3) For triangular, quadrilateral and quad-dominant mixed surface meshes, the MQ always increases and the MSE decreases in increasing iterations. For tri-dominant mixed mesh, the quality of triangles always descends while that of quads ascends.

4) Test examples shows that the MDM is convergent for both planar and surface triangular, quadrilateral and tri-quad meshes.

5) There are some `sharp' vertices in the smoothed surface meshes. One of
the probable cause is that both ridge and corner vertices are strictly
fixed; the other possibly is to project relocated position onto the original
meshes rather than an uniform underlying surface or a local parametric curve
or surface.

\newpage

\begin{acknowledgement}
This research was supported by the Natural Science Foundation of China (Grant Numbers 40602037 and 40872183) and the Fundamental Research Funds for the Central Universities of China.
\end{acknowledgement}

% Non-BibTeX users please follow the syntax
% the syntax of "referenc.tex" for your own citations
%%%%%%%%%%%%%%%%%%%%%%%% referenc.tex %%%%%%%%%%%%%%%%%%%%%%%%%%%%%%
% sample references
% "engineering"
%
% Use this file as a template for your own input.
%
%%%%%%%%%%%%%%%%%%%%%%%% Springer-Verlag %%%%%%%%%%%%%%%%%%%%%%%%%%

%
% BibTeX users please use
% \bibliographystyle{}
% \bibliography{}

%%%%%%%%%%%%%%%%%%%%%%%%%%%%%%%%%%%%%%%%%%%%%%%%%%%%%%%%%%%%%%%%%%%%%%

%%%%%%%%%%%%%%%%%%%%%%%%%%%%%%%%%%%%%%%%%%%%%%%%%%%%%%%%%%%%%%%%%%%%%%

\end{document}